\documentclass[a4paper]{jpconf}
\usepackage{graphicx}
\usepackage{bm,color}

\begin{document}
\title{Evaporation Prescription for \\
Time-Dependent Density Functional Calculations }

\author{Yoritaka Iwata}
\address{GSI Helmholtzzentrum f\"ur Schwerionenforschung, Darmstadt, Germany}
\ead{y.iwata@gsi.de}

\author{Sophia Heinz}
\address{GSI Helmholtzzentrum f\"ur Schwerionenforschung, Darmstadt, Germany}

\begin{abstract}
Collisions between $^{248}$Cm and $^{48}$Ca are systematically calculated by time-dependent density functional calculations with evaporation prescription.
Depending on the incident energy and impact parameter, fusion, fusion-fission, and quasi-fission events are expected to appear.
In this paper, the evaporation prescription is introduced, which is expected to be rather important to heavy-ion reactions producing superheavy nuclei, where the heavier total mass can be related to the higher total excitation energy.
\end{abstract}

\section{Introduction}
\begin{figure}
\begin{center}
\includegraphics[width=8cm]{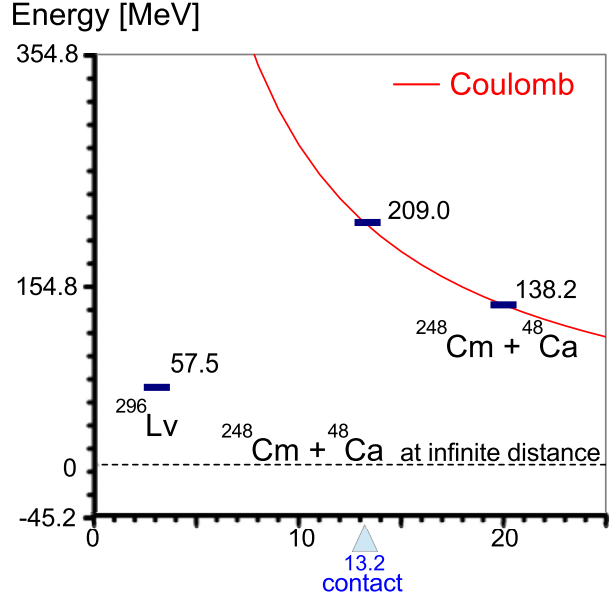}
\end{center}
\caption{\label{fig1a} (Colour online) Potential energy as function of nucleus-nucleus distance $R$ is shown for $^{248}{\rm Cm}+^{48}{\rm Ca}$. 
The binding energies of $^{248}$Cm, $^{48}$Ca and $^{296}{\rm Lv}$ are calculated by static density functional calculations, where the former two nuclei are actually used in TDDFT calculations.
The Coulomb potential is shown by the red curve.}
\end{figure}

The synthesis of superheavy chemical elements~\cite{07hofmann,06oganessian} in the laboratory is one of the biggest challenges in nuclear physics.
It is an attempt for clarifying the existence limits of all the chemical elements, as well as the completion attempt of the periodic table of chemical elements.
We are concerned with heavy-ion collisions
\[  ^{248}{\rm Cm}~+~^{48}{\rm Ca} \]
with different energies and impact parameters in this paper (Fig. \ref{fig1a}).
Let $A$ and $Z$ be the mass number and the proton number, respectively.
The neutron number $N$ is defined by $N = A -Z$, so that $N/Z$ of $^{248}$Cm and $^{48}$Ca are 1.58 and 1.40, respectively.
If fusion appears, $^{296}{\rm Lv}$ (= $^{296}116$) with $N/Z = 1.55$ is produced.

Fast charge equilibration~\cite{10iwata} is expected to appear in low-energy heavy-ion reactions with an incident energy of a few MeV per nucleon.
It provides a very strict limitation for the synthesis of superheavy elements.
Actually, the $N/Z$ of final product is not above nor below the $N/Z$ of the projectile and the target (1.40 $\le$ N/Z $\le$ 1.58 in this case) in the case of charge equilibration, so that the proton-richness of the final product follows.
Although the actual value of $N/Z$ depends on the two colliding ions, its value for the merged nucleus tends to be rather proton-rich for a given proton number of the merged system.
This feature is qualitatively understood by the discrepancy between the $\beta$-stability line and the $N=Z$-line for heavier cases. 

In this paper, based on the time-dependent density functional theory (TDDFT), the reaction dynamics of $^{248}{\rm Cm}+^{48}{\rm Ca }$ is discussed focusing on both, thermal property and charge equilibration dynamics.
Starting with specific features in TDDFT, an implementation of an evaporation prescription in TDDFT is explained (Sec.~\ref{sec1}); the TDDFT calculations are presented for both, before and after the evaporation prescription in Sec.~\ref{sec2}; and the reaction dynamics is summarized in Sec.~\ref{sec3}. 

\section{Identification of evaporation residues} \label{sec1}
\subsection{Treatment of the thermal property}
Self-consistent time-dependent density functional calculations are employed in this paper.
TDDFT reproduces the quantum transportation due to the collective dynamics. 
In this sense, what is calculated by the TDDFT can be regarded as products after several $10^{-21}$~s, which corresponds to a typical time-scale of low-energy heavy-ion reactions (1000~fm/c), as well as to the inclusive time interval of any collective oscillations such as giant dipole resonance, giant quadrupole resonance and so on.
Meanwhile, thermal properties such as the thermal instability are not directly taken into account in TDDFT.
Indeed, the Skyrme type interaction used in TDDFT (for example, see Ref.~\cite{greiner-maruhn}) is determined only from several densities.
It is important that the most effective cooling effect arises from the emission of particles, and therefore it is expected that the break-up or fission of fragments including rather high internal excitation energy is suppressed in the TDDFT final products.
The additional thermal effects leading to the break-ups of fragments should be introduced.

\begin{figure}
\begin{center}
\includegraphics[width=8cm]{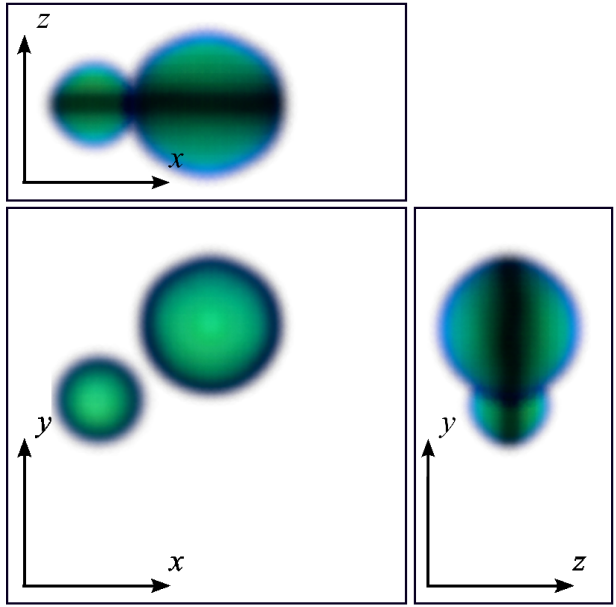}
\end{center}
\caption{\label{fig2a}  (Colour online) Initial configuration for $^{248}{\rm Cm}+^{48}{\rm Ca}$; for instance, a case with $b = 10$~fm is depicted.
Presented box is exactly the same as that used in calculations ($48 \times 48 \times 24$~fm$^3$). }
\end{figure}

Here is a fact that simplifies the treatment of thermal effects, that is, the difference of the time-scales.
Different from the typical time-scale of low-energy heavy-ion reactions, the typical time-scale of the thermal effects is estimated by the typical time interval of collision-fission (fission appearing in heavy-ion collisions): several 10$^{-19}$~s.
It is reasonable to introduce the evaporation prescription simply to the TDDFT final products.
In this context the TDDFT final fragments have the meaning of products just after the early stage of heavy-ion reactions (several $10^{-21}$~s).

\subsection{Evaporation prescription}

In complete fusion reactions the cross-section for the formation of a certain evaporation residue is usually given by three factors \cite{anto}:
\begin{equation}
\sigma_{ER}(E_{cm}) = \sum_J \sigma_{CP} (E_{cm},J) \times P_{CN} (E_{cm},J) 
 \times P_{SV} (E_{cm},J)
\end{equation}
where $\sigma_{CP}$, $P_{CN}$ and $ P_{SV}$ mean the capture cross-section, the probability for the compound nucleus formation, and the probability for survival of the compound nucleus against fission.
All three factors are functions of the centre-of-mass energy $E_{cm}$ and the total angular momentum $J$, where $J$ can be related with the impact parameter. 
For light systems $P_{CN}$ and $P_{SV}$ are about unity and $\sigma_{ER}\,\approx\,\sum_J\sigma_{CP}$. 
But in superheavy systems the strong Coulomb repulsion and large angular momenta lead to small values of $P_{CN}$ and $P_{SV}$ and therefore to the small cross-sections of the evaporation residues observed in the experiments. 
That means, different from light systems, it is necessary to introduce additional thermal effects for the superheavy element synthesis.
First, $\sigma_{CP}$ is sufficiently considered in the TDDFT if we restrict ourselves to a sufficiently high energy exceeding the Coulomb barrier (cf. sub-barrier effects such as tunnelling are not taken into account in the TDDFT).
Second, $P_{CN}$ is fully considered in the TDDFT, which is a kind of mass equilibration also related to charge equilibration.
Third, $P_{SV}$ whose relative time-scale is by no means equal to the former two probabilities is not satisfactorily considered in the TDDFT.
This probability is much more related to thermal effects.
Consequently, further consideration is necessary only for $P_{SV}$ as far as the energy above the Coulomb barrier is concerned.

\begin{figure}
\begin{center}
\includegraphics[width=6cm]{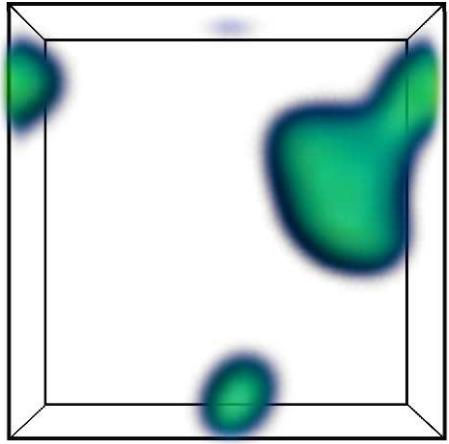}
\end{center}
\caption{\label{fig3a}  (Colour online) Disappearance of charge equilibration for a higher incident energy (1692~MeV), where a case with $b = 10$~fm is shown at 18.6$\times$10$^{-22}$~s.
The charge equilibration upper energy limit is calculated to be $E/A = 3.7$~MeV using the formula shown in Ref.~\cite{10iwata}, and this incident energy is  $E/A = 5.7$~MeV.
$^{19}$O ($N/Z$=1.375) and the fissioning composite nucleus ``$^{277}$108'' are obtained.
Here the production of $^{19}$O whose $N/Z$ is out of the initial range ``$1.40  \le N/Z \le 1.58$'' indicates the disappearance of charge equilibration. 
Note again that the periodic boundary condition is applied here.
}
\end{figure}

Several factors are included in $P_{SV}$ such as probabilities for fission of the compound nucleus,
neutron-evaporation, proton-evaporation, deuteron-evaporation, alpha-particle-evaporation and so on.
As a very first trial of our long-term project of having a generalization of TDDFT calculations, let us take the probability for neutron-evaporation and alpha-particle evaporation:
\[ P_{SV} :=(1- P_{n, evap}) (1- P_{\alpha, evap}) \]
where this choice of neutron and alpha-particle evaporation is based on the experimental insight and results of the present reaction $^{248}{\rm Cm}+^{48}{\rm Ca}$.

The algorithmic description of the evaporation prescription is as follows: \\ \\
1) obtain TDDFT fragments ($A_i$ and $Z_i$) with excitation ($E^*_i$) and kinetic ($K_i$) energies; \\
2) for each fragment calculate evaporation energy: 
\[ \left\{  \begin{array}{ll}
E_{i, \alpha, evap} = E_{i, 1\alpha, evap} + E_{i, \alpha, kin}, \vspace{1.5mm} \\
E_{i, n, evap} = E_{i, 1n, evap} + E_{i, n, kin}, 
\end{array} \right.  \]
where $E_{i, \alpha, evap}$ and $E_{i, n, evap}$ are the cooling energies due to one alpha-particle and one neutron emissions, respectively, and $E_{i, \alpha, kin}$ and $ E_{i, n, kin}$ denotes the kinetic energy of alpha-particle and neutron, respectively (cf. Boltzmann distribution); \\
3) for each fragment find the maximum integer $n_{i,\alpha}$ satisfying
\[ \begin{array}{ll}
 E^*_i \ge n_{i, \alpha} E_{i, \alpha, evap}, 
\end{array}  \]
where $n_{i,\alpha}$ means the maximum number for alpha-particle emission from the $i$-th fragment; \\
4) for each fragment find the maximum integer $n_{i,n}$ satisfying
\[ \begin{array}{ll}
 E^*_i - n_{i, \alpha} E_{i, \alpha, evap} \ge  n_{i, n} E_{i, n, evap}, 
\end{array}  \]
where $n_{i,n}$ means the maximum number for neutron emission from the $i$-th fragment. \\ \\
For the $i$-th fragment, a fragment with mass number $A_i - 4 n_{i, \alpha}- n_{i, n}$ and $Z_i- 2 n_{i, \alpha}$ is produced by emitting $n_{i, \alpha}$ times an alpha-particle and $n_{i, n}$ times a neutron. 
The excitation energy for the fragment is equal to $ E^*_i - n_{i, \alpha} E_{i, \alpha, evap} -  n_{i, n} E_{i, n, evap}$.
Consequently, the total number for alpha-particle and neutron emissions are determined by  $n_{\alpha} = \sum_i n_{i,\alpha}$ and $n_{n} = \sum_i n_{i,n}$, respectively.
Hereafter we assume that neutrons, which are easier to be emitted compared to alpha-particles, are emitted earlier than alpha particles.
This evaporation prescription is preferably implemented directly to the TDDFT results without having the particle number projection or similar statistical treatment for the mass distribution of final fragments.
Indeed, if such a treatment is applied, the excitation energy of the obtained fragments cannot be identified, and shell effects, which are expected to be included in the pure TDDFT results, might be spoiled.

\begin{figure}
\begin{center}
\includegraphics[width=10cm]{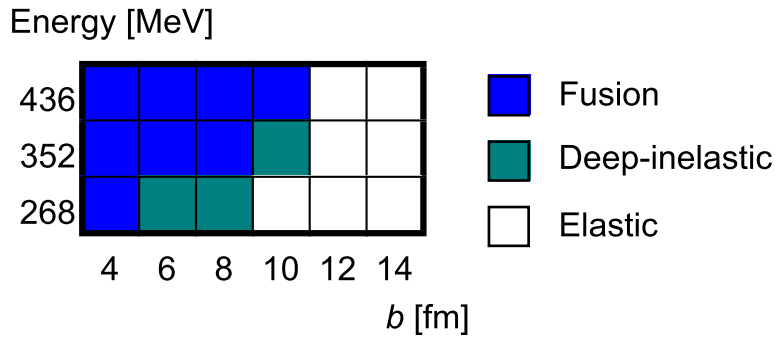}
\end{center}
\caption{\label{fig4a} (Colour online) Diagram of different reaction channels reactions obtained in $^{248}{\rm Cm}+^{48}{\rm Ca}$ collisions by TDDFT.
The preferred reaction channels for different beam energies and impact parameters are given.
The beam energies are all located above the Coulomb barrier which is 209.00~MeV.
The results show fusion, deep-inelastic, and elastic events.}
\end{figure}

\begin{figure}
a) ${\it b} = 8$~fm  \\
\includegraphics[width=3.7cm]{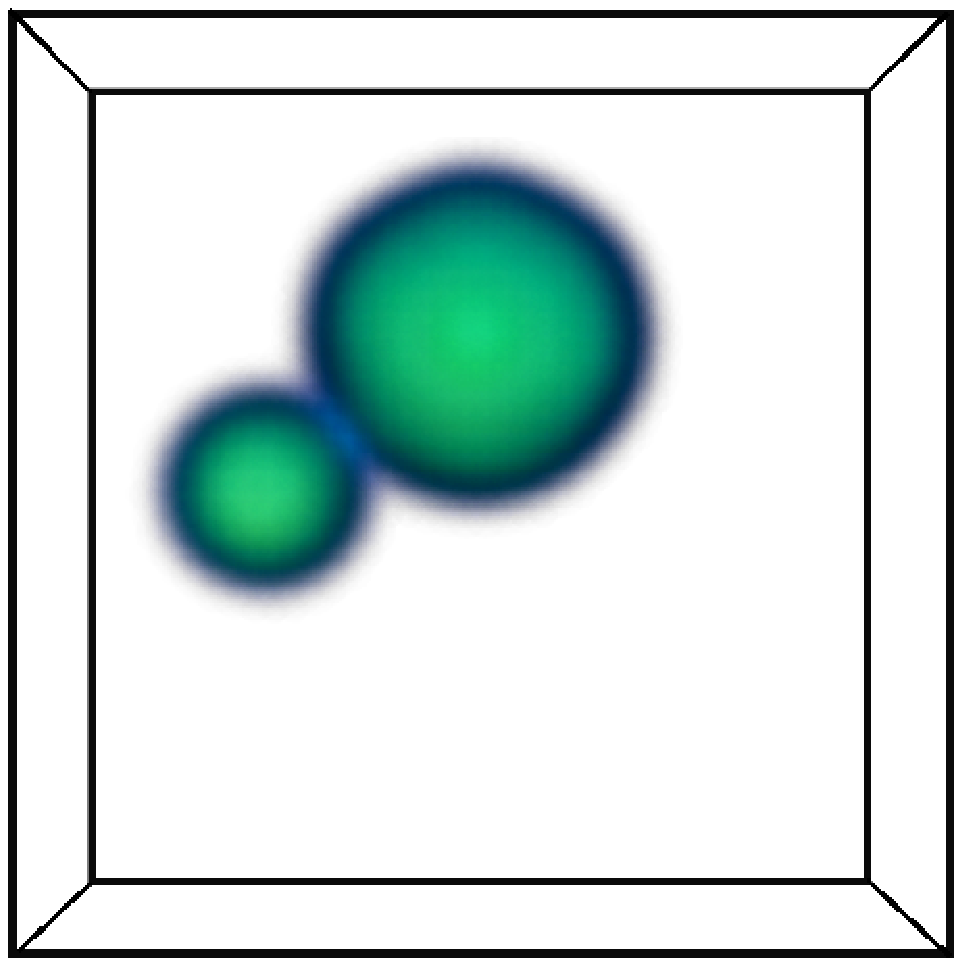}
\includegraphics[width=3.7cm]{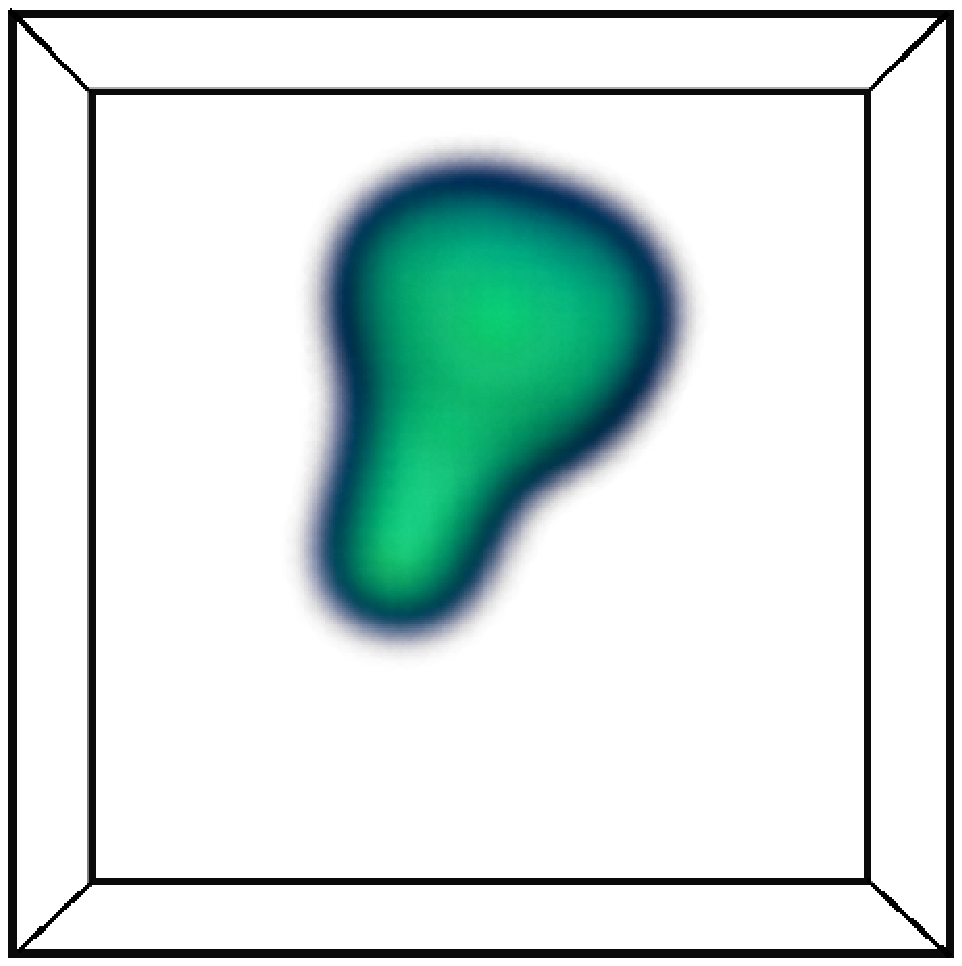}
\includegraphics[width=3.7cm]{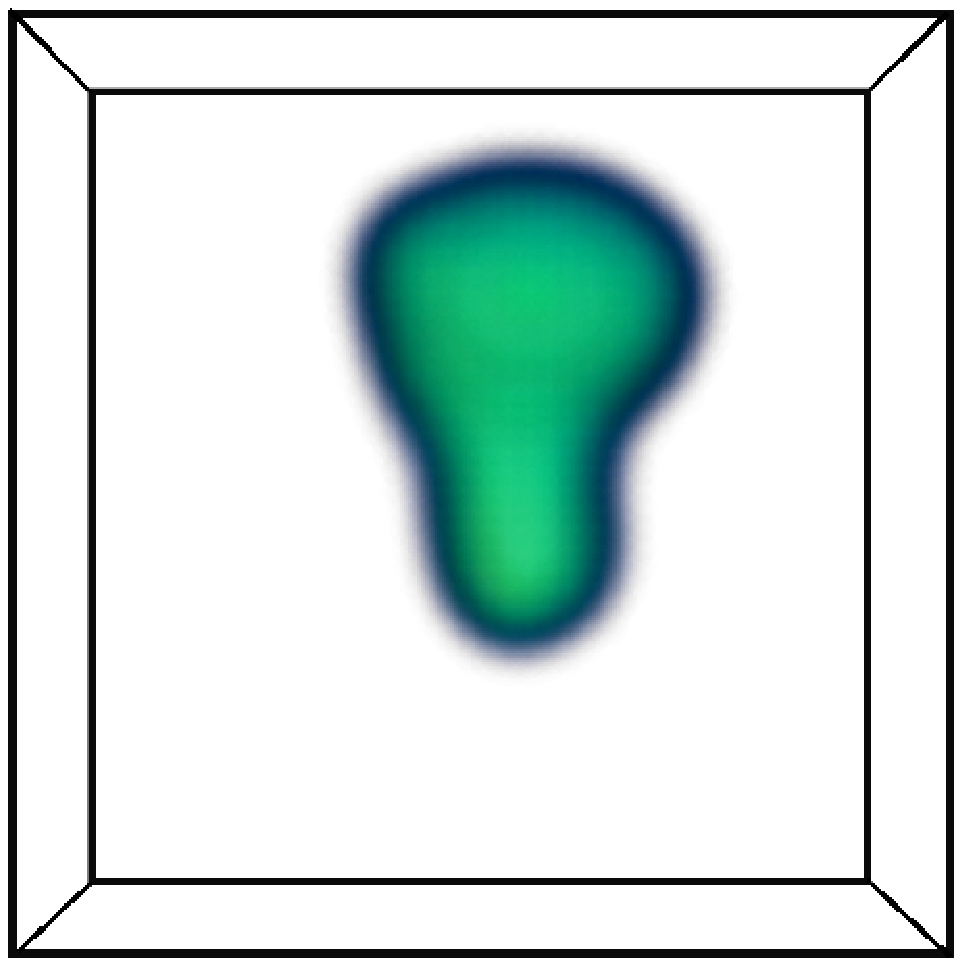}
\includegraphics[width=3.7cm]{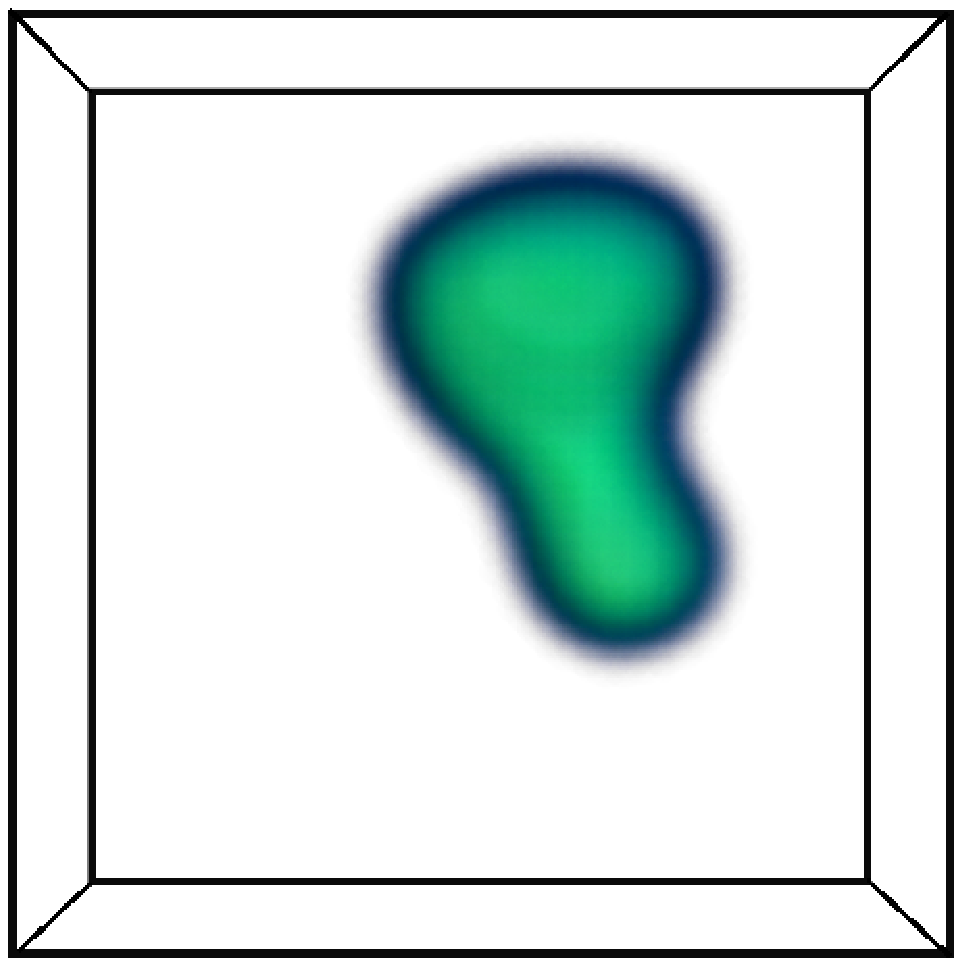}\\
b) ${\it b} = 10$~fm  \\ 
\includegraphics[width=3.7cm]{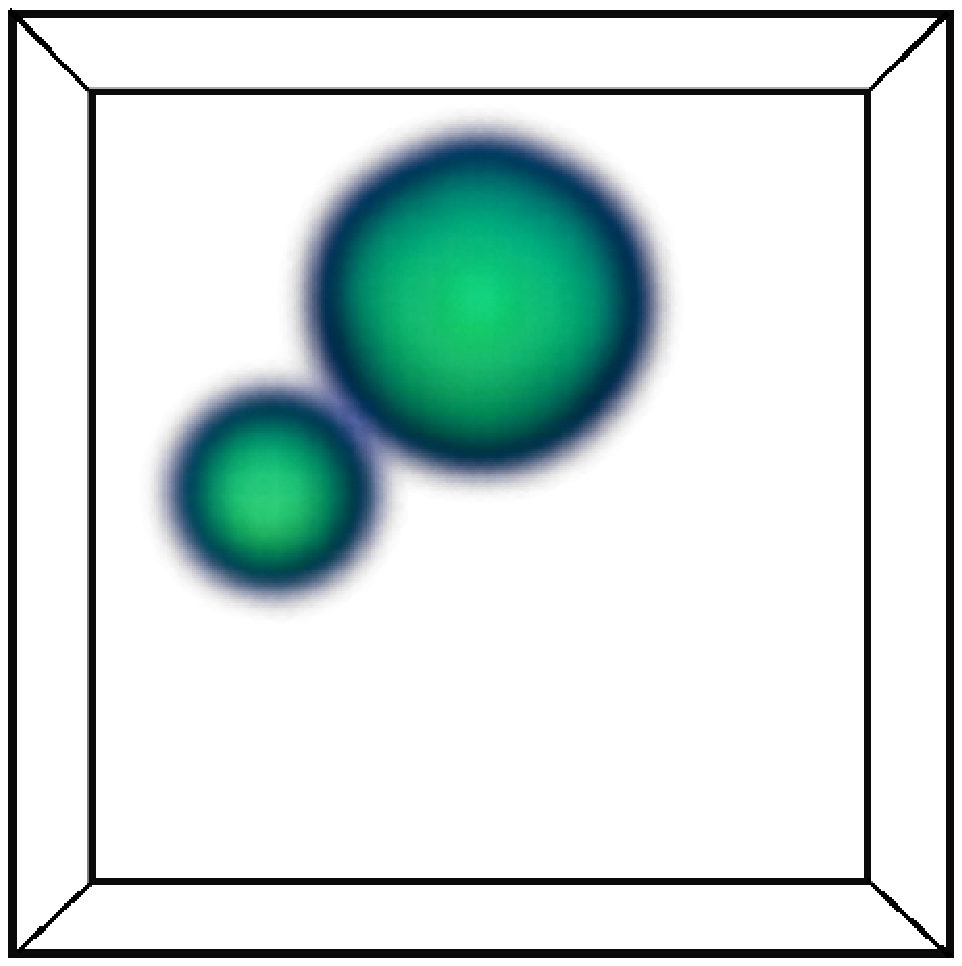}
\includegraphics[width=3.7cm]{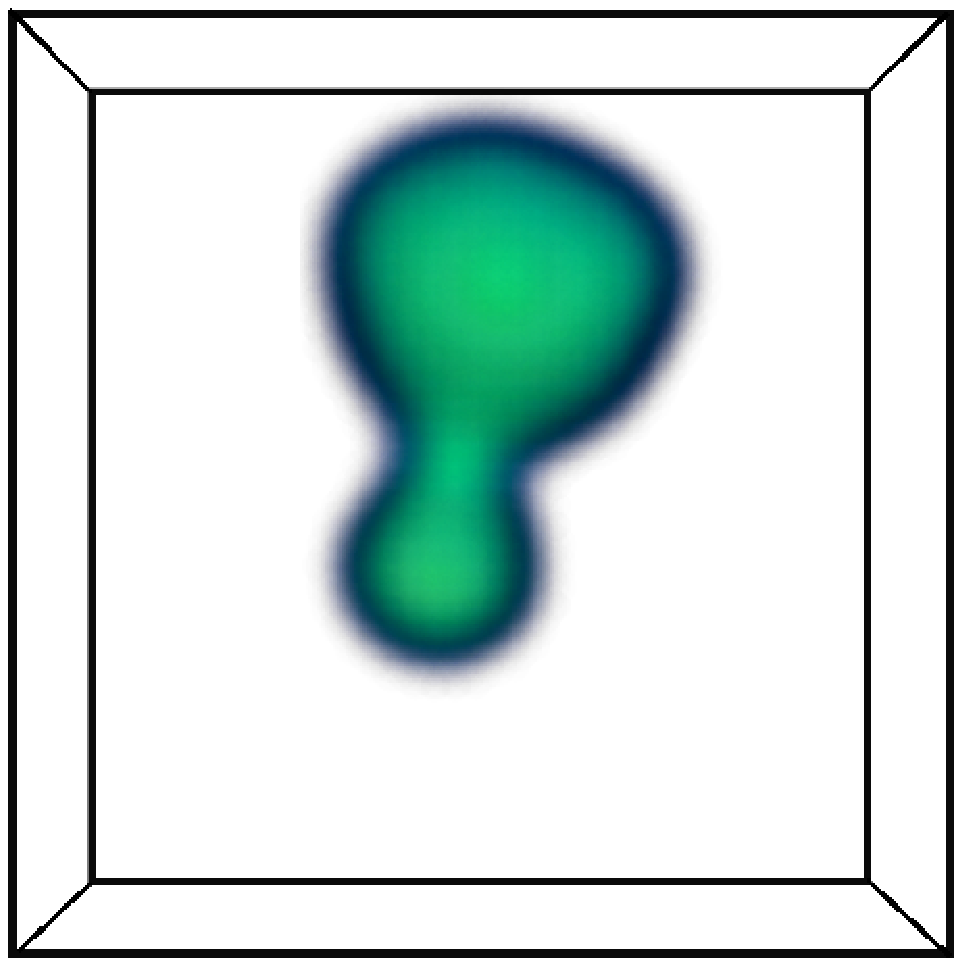}
\includegraphics[width=3.7cm]{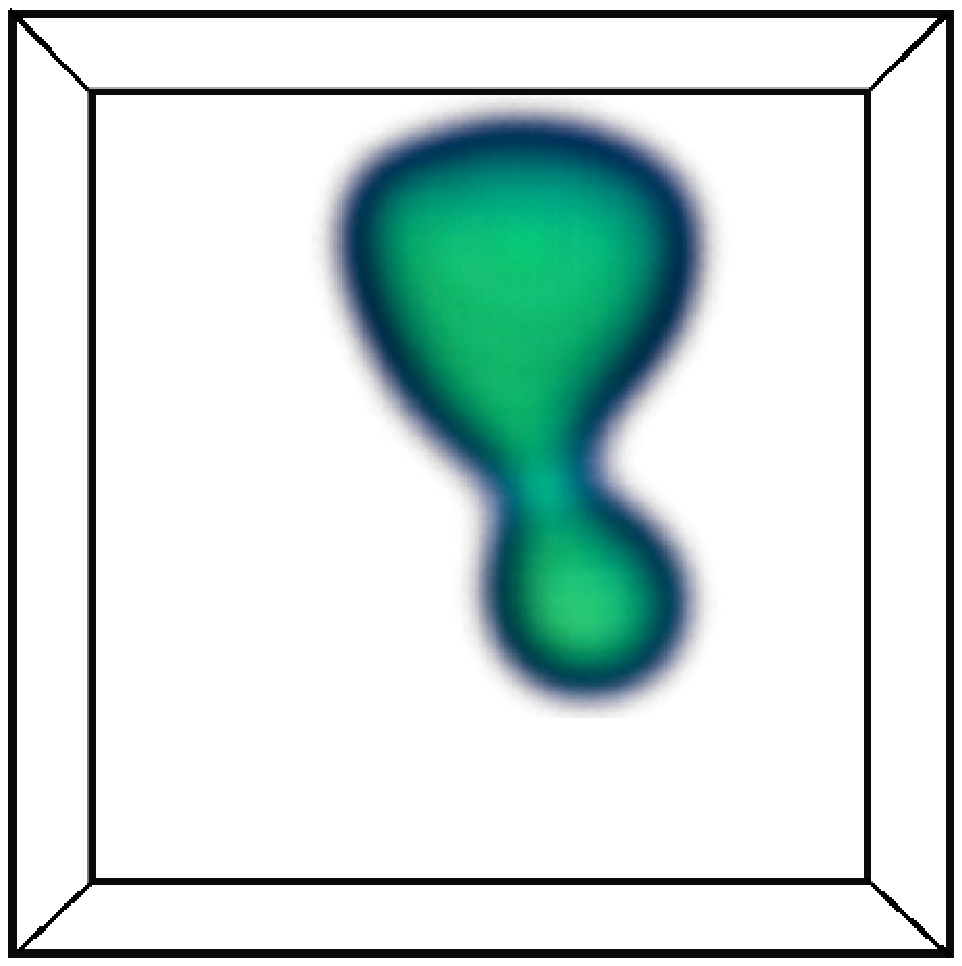}
\includegraphics[width=3.7cm]{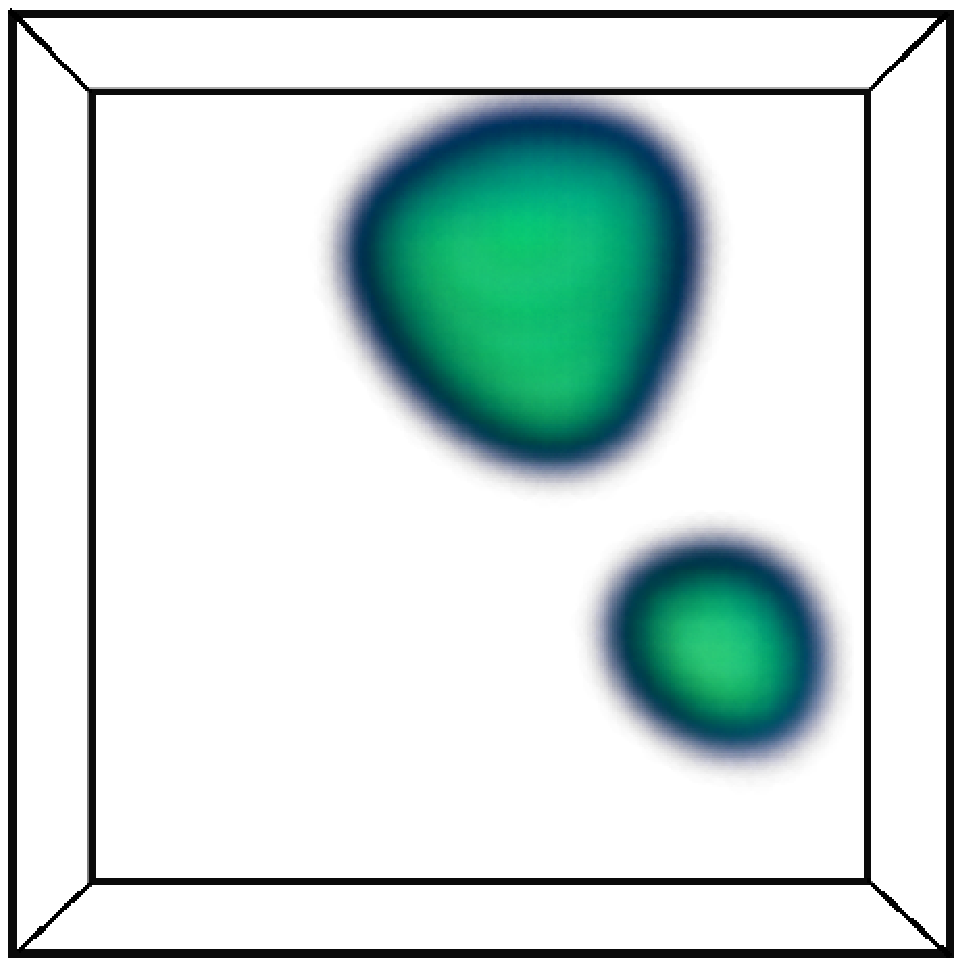}\\
c) ${\it b} = 12$~fm  \\ 
\includegraphics[width=3.7cm]{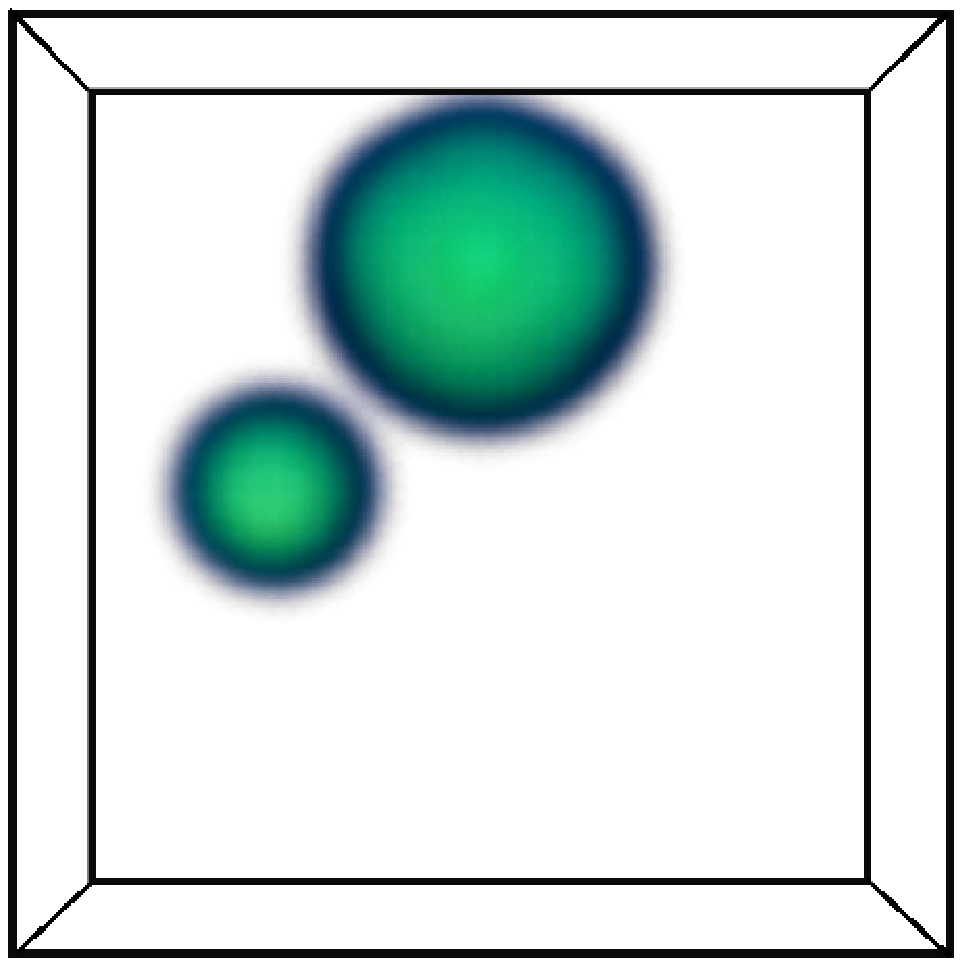}
\includegraphics[width=3.7cm]{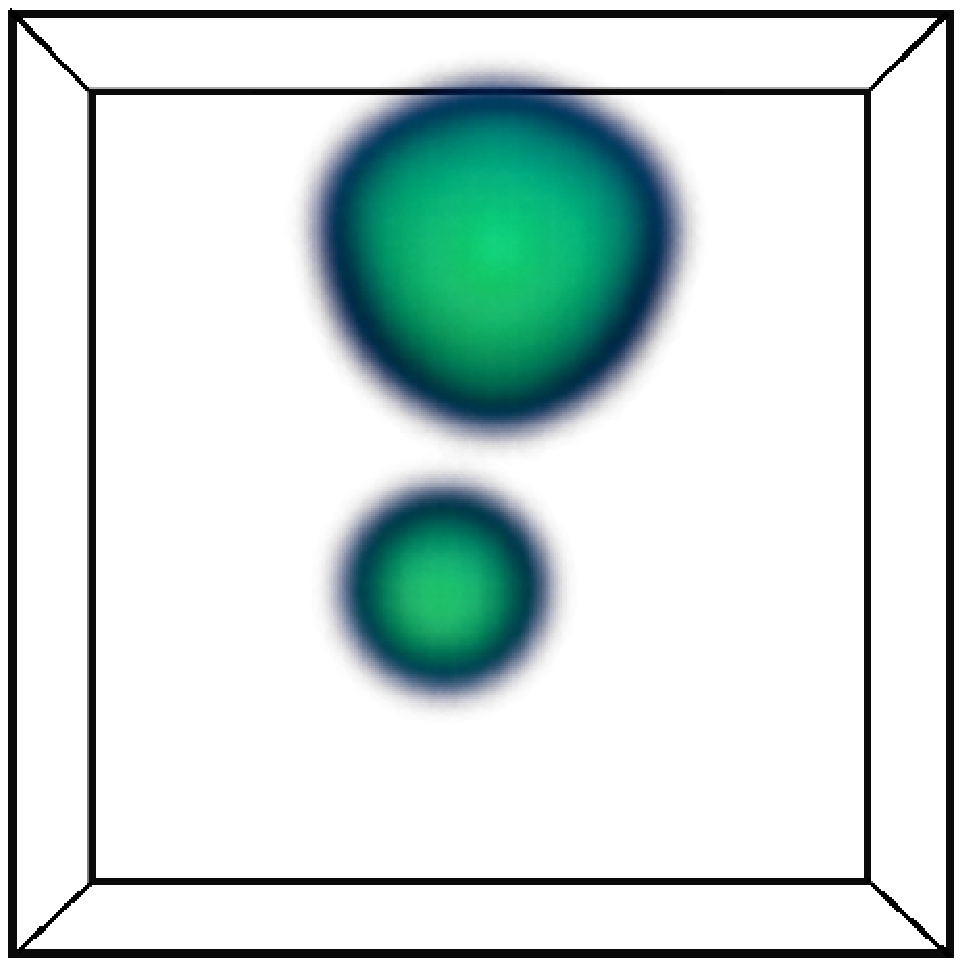}
\includegraphics[width=3.7cm]{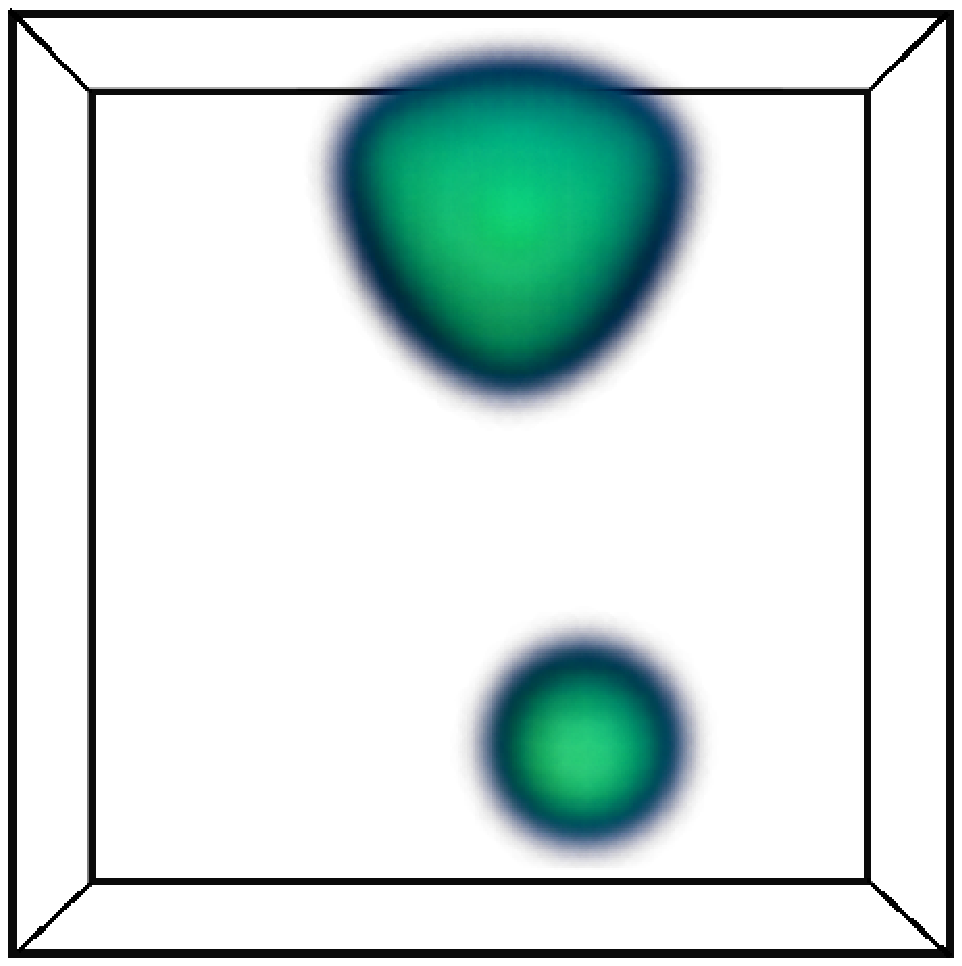}
\includegraphics[width=3.7cm]{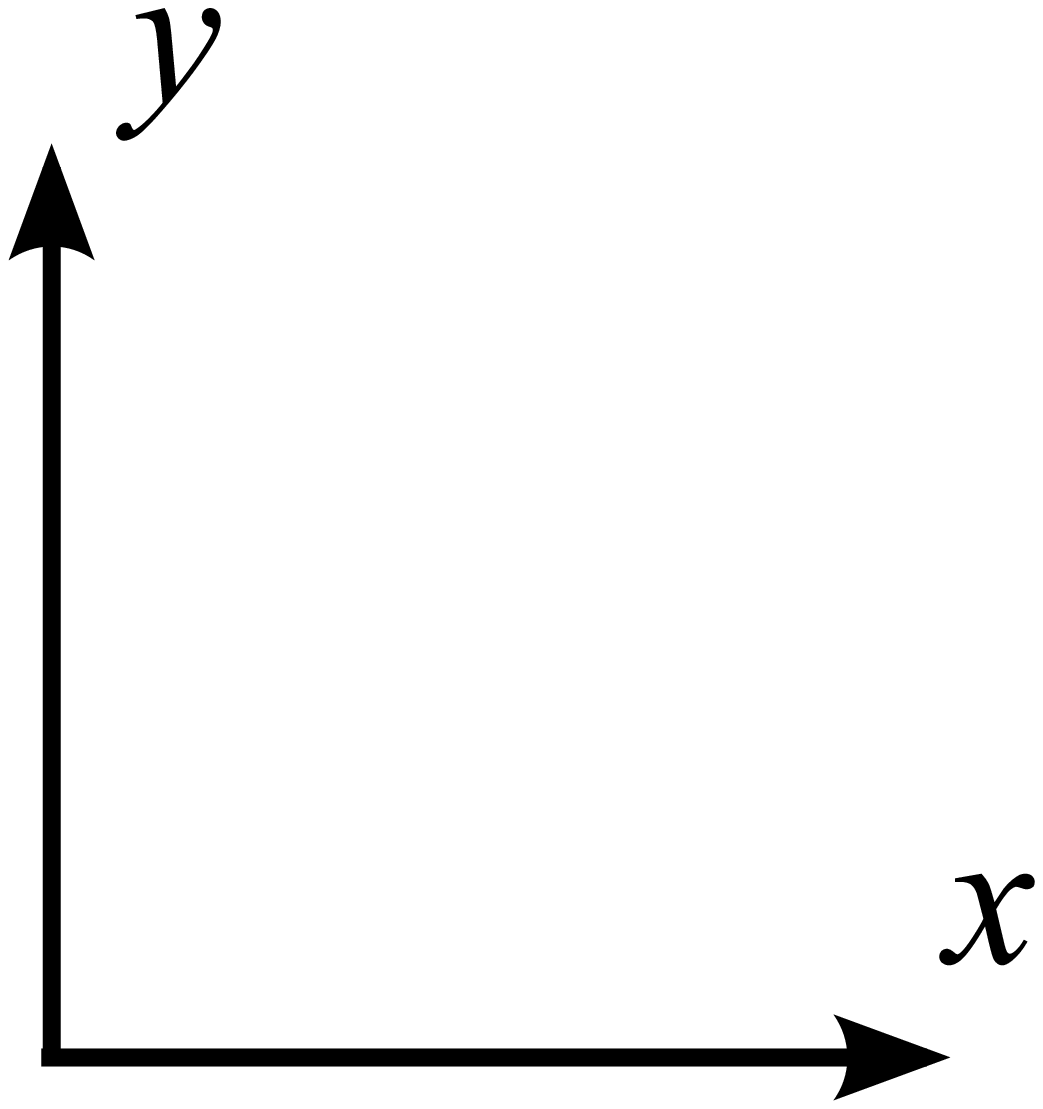}\\
\caption{\label{fig5a} (Colour online) Time evolution of $^{248}{\rm Cm}+^{48}{\rm Ca}$ with a fixed incident energy 352 MeV in the centre-of-mass frame, where a box is fixed to 48$\times$48$\times$24~fm$^3$.
Snapshots at 2.3$\times$10$^{-22}$~s, 8.7$\times$10$^{-22}$~s, 15.1$\times$10$^{-22}$~s and 21.5$\times$10$^{-22}$~s are shown, where a snapshot at 21.5$\times$10$^{-22}$~s is not shown only for ${\it b} = 12$~fm.
Fusion, deep-inelastic, and elastic events appear depending on $b$.
In particular, the life-time of composite nucleus for $b=10$~fm is roughly equal to 15$\times$10$^{-22}$~s.}
\end{figure}

\section{TDDFT calculations for superheavy element synthesis}  \label{sec2}
Time-dependent density functional calculations with a Skyrme interaction (SLy6~\cite{Chabanat-Bonche}) are carried out in a spatial box of $48 \times 48 \times 24$~fm$^3$ with periodic boundary condition.
The unit spatial spacing and the unit time spacing are fixed to 1.0~fm and $2/3 \times 10^{-24}$~s, respectively.

The initial positions of $^{248}$Cm and $^{48}$Ca are fixed to $(0,b,0)$ and (-15,0,0), respectively (Fig. \ref{fig2a}).
The initial $^{248}$Cm is almost spherical; the diameter for $x$, $y$, and $z$ directions are 19~fm, 19~fm and 18~fm, respectively. 
$^{248}{\rm Cm}$ (the right hand side on $x-y$-plane) does not have initial velocity on the frame, while the initial velocity parallel to the $x$-axis is given to $^{48}{\rm Ca}$ (the left hand side on $x-y$-plane).

First of all, the disappearance of charge equilibration for higher energies can be obtained in such reactions producing very heavy nuclei (Fig.~\ref{fig3a}).
This event is a fragmentation without having a compound nucleus, so that it is rather similar to quasi-fission event.
In this paper, the difference between fusion-fission and quasi-fission is identified by whether the fission products satisfy ``1.40 $\le$ N/Z $\le$ 1.58'' or not.
According to the disappearance of charge equilibration, the nucleus with its $N/Z$ satisfying $N/Z<1.40$ or $N/Z>1.58$ can be obtained for higher energies, but the obtained fragments are light.
For the synthesis of neutron-rich and heavy nuclei such as heavy elements located at the island of stability, there is a contradicting situation where high energy (disappearance of charge equilibration) is preferred for neutron-rich synthesis, while low energy (appearance of charge equilibration) is suitable for heavy synthesis.
With respect to producing superheavy elements in the laboratory, here is a reason why we should not go further into higher energies.

The systematic results of TDDFT calculations are summarized in Fig.~\ref{fig4a}.
At a glance, these results, which include many fusion events, provide a quite optimistic view for producing superheavy elements.
However, in comparison with experiments, the corresponding fusion cross-section of those low-energy heavy-ion reactions is too high to believe.
Consequently, although these TDDFT results are still legitimate to show products just after the early stage of heavy-ion reactions, it is necessary to take into account $P_{SV}$ in order to have comparable results to experiments.
 
To compare the difference between the results before and after the evaporation prescription, let us take cases with the incident energy 352~MeV.
TDDFT calculations with different $b$ are shown in Fig.~\ref{fig5a}.
Resulting reactions are 
\begin{equation} \label{reaction-before}  \begin{array}{ll}
 ^{248}{\rm Cm}~+~^{48}{\rm Ca} ~\to~ ^{296}{\rm Lv}  \vspace{1.5mm} \\
 ^{248}{\rm Cm}~+~^{48}{\rm Ca} ~\to~ ^{244}{\rm Cm} +^{52}{\rm Ca}   \vspace{1.5mm} \\
^{248}{\rm Cm}~+~^{48}{\rm Ca} ~\to~ ^{248}{\rm Cm} +^{48}{\rm Ca} 
\end{array} \end{equation}
for $b = 8, 10, 12$~fm, respectively. 
The first, second and third cases correspond to fusion, fission of the composite nucleus (charge equilibration appears), and elastic scattering, respectively.
If the evaporation prescription is introduced to cases with the incident energy 352~MeV, the resulting reactions shown in Eq.~(\ref{reaction-before}) become
\begin{equation} \begin{array}{ll}
 ^{248}{\rm Cm}~+~^{48}{\rm Ca}  ~\to~ ^{275}{\rm Hs} + 4 \alpha + 5 n   \vspace{1.5mm} \\
 ^{248}{\rm Cm}~+~^{48}{\rm Ca}  ~\to~ ^{226}{\rm Th}  +^{50}{\rm Ca}  + 3  \alpha + 8n   \vspace{1.5mm} \\
 ^{248}{\rm Cm}~+~^{48}{\rm Ca}  ~\to~ ^{248}{\rm Cm}  +^{48}{\rm Ca} 
\end{array} \end{equation}
for $b = 8, 10, 12$~fm, respectively.
Alpha-particles and neutrons are emitted for $b \le 10$~fm.
In the same manner, in case of the incident energy 436~MeV (the higher energy), the resulting reaction is
\[ \begin{array}{ll}
 ^{248}{\rm Cm}~+~^{48}{\rm Ca}  ~\to~ ^{264}{\rm Rf} +  6 \alpha + 8 n  
\end{array} \]
for $b \le 10$~fm.
In case of the incident energy 268~MeV (the lower energy), the resulting reactions are
\[ \begin{array}{ll}
 ^{248}{\rm Cm}~+~^{48}{\rm Ca}  ~\to~ ^{280}{\rm Ds}  + 3 \alpha + 4 n   \vspace{1.5mm} \\
 ^{248}{\rm Cm}~+~^{48}{\rm Ca}  ~\to~ ^{238}{\rm Pu}  +^{47}{\rm Ar} + 2 \alpha + 3  n   \vspace{1.5mm} \\
^{248}{\rm Cm}~+~^{48}{\rm Ca} ~\to~ ^{237}{\rm Np}  +^{48}{\rm K} + 2 \alpha + 3 n
\end{array} \]
for $b = 4, 6, 8$~fm, respectively.

\section{Summary}  \label{sec3}
As a result, for low-energy reactions above the Coulomb barrier energy, the obtained reactions are summarized.
\[ \begin{array}{ll}
E = 268 {\rm MeV},   \vspace{1.5mm} \\
\qquad  ^{248}{\rm Cm}+^{48}{\rm Ca}  ~\to~ ^{292}{\rm Lv} + 4 n   \vspace{1.5mm} \\
\qquad \quad  ~\to~ ^{288}{\rm Fl}  + \alpha + 4 n   \vspace{1.5mm} \\
\qquad \quad  ~\to~ ^{284}{\rm Cn}  + 2 \alpha + 4 n   \vspace{1.5mm} \\
\qquad \quad  ~\to~ ^{280}{\rm Ds}  + 3 \alpha + 4 n,  \quad (b \le 4~{\rm fm})   \vspace{1.5mm} \\
\qquad  ^{248}{\rm Cm}+^{48}{\rm Ca}  ~\to~ ^{246}{\rm Cf}  +^{47}{\rm Ar} + 3  n  \vspace{1.5mm} \\
\qquad \quad  ~\to~ ^{242}{\rm Cm}  +^{47}{\rm Ar}  + \alpha + 3  n  \vspace{1.5mm} \\
\qquad \quad  ~\to~ ^{238}{\rm Pu}  +^{47}{\rm Ar}  + 2 \alpha + 3  n,  \quad (b = 6~{\rm fm})    \vspace{1.5mm} \\
\qquad  ^{248}{\rm Cm}+^{48}{\rm Ca} ~\to~ ^{245}{\rm Bk}  + ^{48}{\rm K}  + 3  n  \vspace{1.5mm} \\
\qquad \quad ~\to~ ^{241}{\rm Am}  + ^{48}{\rm K}  + 1 \alpha + 3  n  \vspace{1.5mm} \\
\qquad \quad ~\to~ ^{237}{\rm Np}  + ^{48}{\rm K}  + 2 \alpha + 3  n,  \quad (b = 8~{\rm fm})     \vspace{1.5mm} \\
\qquad  ^{248}{\rm Cm}+^{48}{\rm Ca} ~\to~ ^{248}{\rm Cm}  +^{48}{\rm Ca}, \quad (b \ge 10~{\rm fm});      \vspace{2.5mm} \\
\end{array} \]
\[ \begin{array}{ll}
E = 352 {\rm MeV},   \vspace{1.5mm} \\
\qquad ^{248}{\rm Cm}+^{48}{\rm Ca}    ~\to~ ^{291}{\rm Lv} + 5n  \vspace{1.5mm} \\
\qquad \quad ~\to~ ^{287}{\rm Fl} + \alpha + 5n  \vspace{1.5mm} \\ 
\qquad \quad ~\to~ ^{283}{\rm Cn} +2 \alpha + 5n  \vspace{1.5mm} \\
\qquad \quad ~\to~ ^{279}{\rm Ds} +3 \alpha + 5n  \vspace{1.5mm} \\
\qquad \quad ~\to~ ^{275}{\rm Hs} +4 \alpha + 5n, \quad (b \le 8~{\rm fm}),  \vspace{1.5mm} \\
\qquad ^{248}{\rm Cm}+^{48}{\rm Ca}  ~\to~ ^{238}{\rm Cm}  +^{50}{\rm Ca}  + 8n   \vspace{1.5mm} \\ 
\qquad \quad ~\to~ ^{234}{\rm Pu}  +^{50}{\rm Ca}  + \alpha + 8n   \vspace{1.5mm} \\ 
\qquad \quad ~\to~ ^{230}{\rm U}  +^{50}{\rm Ca}  + 2 \alpha + 8n  \vspace{1.5mm} \\ 
\qquad \quad ~\to~ ^{226}{\rm Th}  +^{50}{\rm Ca}  + 3 \alpha + 8n, \quad (b = 10~{\rm fm}),   \vspace{1.5mm} \\
\qquad ^{248}{\rm Cm}+^{48}{\rm Ca} ~\to~ ^{248}{\rm Cm}  +^{48}{\rm Ca}, \quad (b \ge 12~{\rm fm});    \vspace{2.5mm} \\
\end{array} \]
\[ \begin{array}{ll}
E = 436 {\rm MeV},   \vspace{1.5mm} \\
\qquad ^{248}{\rm Cm}+^{48}{\rm Ca}  ~\to~ ^{288}{\rm Lv} + 8 $n$  \vspace{1.5mm} \\
\qquad \quad ~\to~ ^{284}{\rm Fl} + \alpha + 8 $n$   \vspace{1.5mm} \\
\qquad \quad ~\to~ ^{280}{\rm Cn} + 2\alpha + 8 $n$  \vspace{1.5mm} \\
\qquad \quad ~\to~ ^{276}{\rm Ds} + 3\alpha + 8 $n$  \vspace{1.5mm} \\
\qquad \quad ~\to~ ^{272}{\rm Hs} + 4\alpha + 8 $n$  \vspace{1.5mm} \\
\qquad \quad ~\to~ ^{268}{\rm Sg} + 5\alpha + 8 $n$  \vspace{1.5mm} \\
\qquad \quad ~\to~ ^{264}{\rm Rf} + 6\alpha + 8 $n$, \quad (b \le 10~{\rm fm}),    \vspace{1.5mm} \\
\qquad ^{248}{\rm Cm}+ ^{48}{\rm Ca} ~\to~  ^{248}{\rm Cm}+ ^{48}{\rm Ca}, \quad (b \ge 12~{\rm fm}),   
\end{array} \]
where the indicated energy is the centre-of-mass energy and $b$ is incremented by 2~fm.
Transfer of a few nucleons contributing to charge equilibration can be noticed for $E$ = 268 and 352~MeV cases.
The experimental results reveal that fission of the compound nucleus occurs with very large probability, which will be included in future steps into the calculations.

\ack{This work was supported by the Helmholtz alliance HA216/EMMI.
The authors thank Prof. J. A. Maruhn for reading this manuscript carefully.}

\end{document}